\documentclass[sigconf]{acmart}

\usepackage[toc]{glossaries}
\usepackage{tabularx,bytefield,csquotes,makecell}
\usepackage[textsize=tiny]{todonotes}
\usepackage{listings}

\usetikzlibrary{backgrounds, fit}
\definecolor{maincolor}{RGB}{170, 132, 57}
\definecolor{firstside}{RGB}{64, 48, 117}
\definecolor{secondside}{RGB}{34, 102, 102}

\AtBeginDocument{%
  }

\copyrightyear{2022} 
\acmYear{2022} 
\setcopyright{acmlicensed}\acmConference[CSCS '22]{Computer Science in Cars Symposium}{December 8, 2022}{Ingolstadt, Germany}
\acmBooktitle{Computer Science in Cars Symposium (CSCS '22), December 8, 2022, Ingolstadt, Germany}
\acmPrice{15.00}
\acmDOI{10.1145/3568160.3570229}
\acmISBN{978-1-4503-9786-5/22/12}

\newacronym[]{can}{CAN}{Controller Area Network}
\newacronym[]{cll}{CLL}{Cryptographic Link Layer}
\newacronym[]{dtc}{DTC}{Diagnostic Trouble Codes}
\newacronym[]{dtls}{DTLS}{Datagram Transport Layer Security}
\newacronym[]{dds}{DDS}{Data Distribution Service}
\newacronym[]{ip}{IP}{Internet Protocol}
\newacronym[]{ipsec}{IPsec}{Internet Protocol Security}
\newacronym[]{lan}{LAN}{Local Area Network}
\newacronym[]{lin}{LIN}{Local Interconnected Network}
\newacronym[]{mitm}{MITM}{man-in-the-middle}
\newacronym[]{most}{MOST}{media oriented systems transport}
\newacronym[]{macsec}{MACsec}{Media Access Control Security}
\newacronym[]{osi}{OSI}{Open Systems Interconnection}
\newacronym[]{pad}{PAD}{peer authentication database}
\newacronym[]{uds}{UDS}{Unified Diagnostic Services}
\newacronym[]{vin}{VIN}{Vehicle Identification Number}
\newacronym[]{gid}{GID}{Group Identifier}
\newacronym[]{eid}{EID}{Entity Identifier}
\newacronym[]{vlan}{VLAN}{Virtual Local Area Network}
\newacronym[longplural={electronic control units}]{ecu}{ECU}{electronic control unit}
\newacronym{doip}{DoIP}{Diagnostic communication over Internet Protocol}
\newacronym{tls}{TLS}{Transport Layer Security}
\newacronym[longplural={source addresses}]{sa}{SA}{source address}
\newacronym[]{secoc}{SecOC}{Secure Onboard Communication}
\newacronym{tcp}{TCP}{Transmission Control Protocol}
\newacronym[]{udp}{UDP}{User Datagram Protocol}
\newacronym[]{pki}{PKI}{public key infrastructure}

\lstdefinelanguage{Tamarin}{
  keywords={typeof, new, true, false, catch, function, return, null, catch, switch, var, if, in, while, do, else, case, break, const, begin, end},
  ndkeywords={class, export, boolean, throw, implements, import, this, theory, builtins, rule, lemma},
  basicstyle=\footnotesize\ttfamily,
  numbers=left,
  numberstyle=\scriptsize,
  stepnumber=1,
  numbersep=8pt,
  showstringspaces=false,
  breaklines=true,
  belowcaptionskip=.75\baselineskip,
  comment=[l]{//},
  morecomment=[s]{/*}{*/},
  morestring=[b]',
  morestring=[b]"
}

\begin{document}
\lstset{language=Tamarin}

%%
%% The "title" command has an optional parameter,
%% allowing the author to define a "short title" to be used in page headers.
\title{Analysis of the DoIP Protocol for Security Vulnerabilities}

%%
%% The "author" command and its associated commands are used to define
%% the authors and their affiliations.
%% Of note is the shared affiliation of the first two authors, and the
%% "authornote" and "authornotemark" commands
%% used to denote shared contribution to the research.

\author{Patrick Wachter}
% \authornote{Both authors contributed equally to this research.}
% \email{patrick.wachter@uni-ulm.de}
\email{patrick.wachter@mercedes-benz.com}
\orcid{0000-0001-9329-640X}
% \author{P. Wachter}
%\authornotemark[1]
\affiliation{%
	\institution{Mercedes-Benz Tech Innovation GmbH}
	\streetaddress{Wilhelm-Runge-Straße 11}
	\city{Ulm}
	% \state{Ohio}
	\country{Germany}
	\postcode{89081}
}

\author{Stephan Kleber}
%% \authornote{Both authors contributed equally to this research.}
%% \email{stephan.kleber@uni-ulm.de}
\email{stephan.kleber@mercedes-benz.com}
\orcid{0000-0001-9836-4897}
% \author{S. Kleber}
%\authornotemark[1]
\affiliation{%
  \institution{Mercedes-Benz Tech Innovation GmbH}
  \streetaddress{Wilhelm-Runge-Straße 11}
  \city{Ulm}
  % \state{Ohio}
  \country{Germany}
  \postcode{89081}
}

%%
%% By default, the full list of authors will be used in the page
%% headers. Often, this list is too long, and will overlap
%% other information printed in the page headers. This command allows
%% the author to define a more concise list
%% of authors' names for this purpose.
%\renewcommand{\shortauthors}{Wachter and Kleber}

%%
%% The abstract is a short summary of the work to be presented in the
%% article.
\begin{abstract}
  DoIP, which is defined in ISO\,13400, is a transport protocol stack for diagnostic data. Diagnostic data is a potential attack vector at vehicles, so secure transmission must be guaranteed to protect sensitive data and the vehicle. 
  Previous work analyzed a draft version and earlier versions of the DoIP protocol without \gls*{tls}.
  No formal analysis exists for the DoIP protocol.

  The goal of this work is to investigate the DoIP protocol for design flaws that may lead to security vulnerabilities and possible attacks to exploit them.
  For this purpose, we deductively analyze the DoIP protocol in a first step and subsequently confirm our conclusions formally.
  For the formal analysis, we use the prover Tamarin.
  Based on the results, we propose countermeasures to improve the DoIP's security.
  We show that the DoIP protocol cannot be considered secure mainly because the security mechanisms \gls*{tls} and client authentication in the DoIP protocol are not mandatory.
  We propose measures to mitigate the vulnerabilities that we confirm to remain after activating \gls*{tls}.
  These require only a minor redesign of the protocol.
\end{abstract}

% The code below is generated by the tool at http://dl.acm.org/ccs.cfm.
% Please copy and paste the code instead of the example below.
%


%% TODO: Add keywords, uncomment line below.
%% Keywords. The author(s) should pick words that accurately describe
%% the work being presented. Separate the keywords with commas.
%% A "teaser" image appears between the author and affiliation
%% information and the body of the document, and typically spans the
%% page.

%% TODO: Add a teaser image if we find a suitable one.

% \begin{teaserfigure}
%   \includegraphics[width=\textwidth]{sampleteaser}
%   \caption{Seattle Mariners at Spring Training, 2010.}
%   \Description{Enjoying the baseball game from the third-base
%   seats. Ichiro Suzuki preparing to bat.}
%   \label{fig:teaser}
% \end{teaserfigure}

%%
%% This command processes the author and affiliation and title
%% information and builds the first part of the formatted document.
\maketitle

\section{Introduction}
% !TeX spellcheck = en_US
Modern cars use more than 70 \glspl*{ecu} and this number will continue to rise~\cite{staron_automotive_2021}. 
When the number of \glspl*{ecu} in cars was lower, it was relatively easy to maintain and update them. 
In the workshop, with a tester directly connected to the car via dedicated wiring, the diagnostic data was read and minor updates were installed. 
However, the requirements for modern cars changed and nowadays it is desired to read out diagnostic data from anywhere via remote diagnostics. 
Over-the-air updates must also be possible~\cite{ajin_study_2016}.
%Artificial intelligence also plays a major role with the large amount of real world data. 
Due to the ever-increasing number of \glspl*{ecu} and their increasing complexity, such updates are becoming larger, in some cases several gigabytes. 
This is no longer feasible with conventional bus systems such as \gls*{can}. 
This is why Ethernet and the \gls*{ip} were adopted in the car.
As a transport protocol the \enquote{\gls*{doip}} is used in this context.
\gls*{uds}, which is traditionally used over \gls*{can}, is the application protocol also used with \gls*{doip}~\cite{iso_iso_2011, iso_iso_2020, iso_iso_2013}. 
The diagnostic interface allows access to information of the \glspl*{ecu}. 
It can be used to retrieve emission-related diagnostic data or to perform manufacturer-specific functions, such as calibrating technical components and software updates~\cite{iso_iso_2011}. 
However, the diagnostic interface of cars can also be used by attackers to gain access to the vehicle's electronic system.

In this paper, we examine the security of \gls*{doip} and perform a formal analysis of individual protocol phases.
We thus provide a thorough assessment of the design's robustness against malicious action that could undermine a vehicle's functionality and safety.

\section{Background and Related Work}
% !TeX spellcheck = en_US
%% TODO: Remove the \medskip and other skips.

\gls*{doip} is defined in ISO\,13400~\cite{iso_iso_2011,iso_iso_2019,iso_iso_2016,iso_iso_2016-1} for the usage with the \gls*{udp} for broadcasting initial messages and with the \gls*{tcp} after a session is established. 
\gls*{doip} covers aspects of the layers one to four of the \gls*{osi} basic reference model~\cite{iso_isoiec_1994} as illustrated in \autoref{fig:osi-doip}. 
Also, the usage of Ethernet, \gls*{ip} and some auxiliary protocols is defined in the specification~\cite{iso_iso_2011}.
In the latest version of the standard, the use of \gls*{tls} has also been added~\cite{iso_iso_2019}. 
%On the level of \gls*{doip}, there are no authentication mechanisms intended. Authentication is only optional and is recommended on the application layer~\cite{iso_iso_2019}. This means that wide varieties of attacks are possible.

\begin{figure}
    \begin{tikzpicture}[xscale=.5, yscale=.5]
        \sffamily
        \node[anchor=base] at (0,7) (layer7) {7};
        \node[anchor=base] at (0,6) (layer6) {6};
        \node[anchor=base] at (0,5) (layer5) {5};
        \node[anchor=base] at (0,4) (layer4) {4};
        \node[anchor=base] at (0,3) (layer3) {3};
        \node[anchor=base] at (0,2) (layer2) {2};
        \node[anchor=base] at (0,1) (layer1) {1};
        \node[anchor=base west] at (.6,7) (text7) {Application Layer };
        \node[anchor=base west] at (.6,6) (text6) {Presentation Layer};
        \node[anchor=base west] at (.6,5) (text5) {Session Layer     };
        \node[anchor=base west] at (.6,4) (text4) {Transport Layer   };
        \node[anchor=base west] at (.6,3) (text3) {Network Layer     };
        \node[anchor=base west] at (.6,2) (text2) {Data Link Layer   };
        \node[anchor=base west] at (.6,1) (text1) {Physical Layer    };

        \node[anchor=base] at (8,7) (tcpip7) {\textbf{UDS} on IP};        
        \node[anchor=base] at (8,5) (tcpip5) {\textbf{UDS} Session};        
        \node[anchor=base] at (8,4) (tcpip4) {TCP/UDP   };
        \node[anchor=base] at (8,3) (tcpip3) {IPv4/IPv6 };
        \node[anchor=base] at (8,2) (tcpip2) {Ethernet  };
        \node[anchor=base] at (8,1) (tcpip1) {IEEE\,802.3};
        
        \node[anchor=base east, fill=firstside, inner sep=1pt, text=white] at (15.5,7) (doip7) {\textbf{Vehicle Discovery}};
        \node[anchor=base east, fill=firstside, inner sep=1pt, text=white] at (15.5,5) (doip5) {\textbf{DoIP Header\vphantom{y}}      };
        \node[anchor=base east, fill=firstside, minimum height=2.5em, text=white] at (15.5,3.5) (doip3) {\textbf{ISO\,13400--2}};
        \node[anchor=base east, fill=secondside, minimum height=2.5em, text=white] at (15.5,1.5) (doip1) {\textbf{ISO\,13400--3}};
        
        \begin{scope}[on background layer, every node/.style={fill=maincolor!60, inner sep=0}]
            \foreach \x in {layer7,layer6,layer5,layer4,layer3,layer2,layer1} {
                \fill[maincolor!40] (\x.north west) -- +(16,0) |- (\x.south west) -- cycle;
            }
        \end{scope}
    \end{tikzpicture}
	\caption{\gls*{osi} layer model related to \gls*{doip} parts.}
	\label{fig:osi-doip}
\end{figure}

In 2011, \citeauthor{lindberg_security_2011} was the first to examine a late draft version of the \gls*{doip} protocol~\cite{lindberg_security_2011} for its security properties. 
When he authored his thesis in 2011, \gls*{doip} had not yet officially been released, so there is an urgent need to revisit and investigate the protocol again.
\citeauthor{lindberg_security_2011} concluded that \gls*{doip} as of 2011 is not a secure protocol without additional measures. 
His findings reveal that one of the biggest vulnerabilities is the lack of authentication and of ensuring data integrity.
He also notices ambiguities in the specification and a missing plausibility check for the validation of data in received messages~\cite{lindberg_security_2011}.

Published in 2016, \citeauthor{ajin_study_2016} compared different technologies in vehicle networks and gave a security analysis of \gls*{doip}~\cite{ajin_study_2016}.
\citeauthor{ajin_study_2016} came to the same conclusion as \citeauthor{lindberg_security_2011}: Various security vulnerabilities exist in \gls*{doip}. In 2016, \gls*{tls} was not yet present in the \gls*{doip} protocol specification.

\citeauthor{matsubayashi_masaru_attacks_2021} evaluated attacks against \gls*{uds} on \gls*{doip}~\cite{matsubayashi_masaru_attacks_2021}.
They used a previous version of the \gls*{doip} protocol in which \gls*{tls} was not yet present and found three attacks through the analysis: The first one is authentication avoidance in \gls*{uds}.
The other two attacks are \gls*{mitm} attacks by session hijacking through vehicle announcement messages and vehicle identification responses.
They have demonstrated these attacks in an experimental environment and presented countermeasures for them~\cite{matsubayashi_masaru_attacks_2021}.

\citeauthor{kleberger_securing_2014} addressed the issue of how diagnostic data can be transferred securely, and how the diagnostic infrastructure in workshops can be secured~\cite{kleberger_securing_2014}.
To accomplish this, they compared four different options and concluded that \gls*{ipsec} is the approach to take.
It is easy to deploy and to maintain, although tools would have to be developed to help manage the \gls*{pad}.
\gls*{ipsec} would also secure remote vehicle diagnostics over the internet.
The key agreement protocol IKE can be replaced by a protocol specially designed for the use case, which distributes session keys and security policies only to authorized equipment~\cite{kleberger_securing_2014}.
The paper was written before \gls*{tls} was part of \gls*{doip}.

Most related work on \gls*{doip} security discusses the versions without \gls*{tls}.
Its addition requires revisiting their findings and may render solutions like \gls*{ipsec} obsolete.
Our paper analyzes the current protocol version and addresses the need for an updated security assessment.

\citeauthor{lauser_security_2020} examined the status of the analysis of automotive vehicle protocols. 
Therefore, they give an overview of tools for formal analysis and, as an example, analyze AUTOSAR's \gls*{secoc} with Tamarin~\cite{lauser_security_2020}.
As future work, they propose a formal analysis of the \gls*{doip} protocol and, especially, \gls*{doip} with \gls*{tls}.
They conclude that formal analysis with a symbolic model is well suited for automotive protocols:
Suitable tools exist that provide a quick and easy way to analyze automotive protocols~\cite{lauser_security_2020}.

\subsection{Unified Diagnostic Services and Use Cases}
The \gls*{uds} protocol is defined in ISO standard 14229 and is used for diagnostic services requests and replies.
A tester connected to the vehicle uses \gls*{uds} in a client-server relationship with the tester being the client and the vehicle being the server~\cite{iso_iso_2020}.

Connecting a tester allows different use cases.
During the production of the vehicle, \gls*{doip} is used to activate the software of the vehicle, to correct errors via software, or to detect hardware failures like broken sensors.
During the vehicle lifetime, it is possible to read \gls*{dtc} to find errors and clear them. 
It is also possible to read various parameters of the car, including the temperature, the state of charge, or the \gls*{vin}. In addition, diagnostic sessions can be initiated: Among other things, this allows the testing of safety-relevant features. The last use case is the modification of the \gls*{ecu} behavior: This is possible through resets, firmware flashing, or changes of settings~\cite{iso_iso_2020}.
Since \gls*{uds} per se has no security protection for payload data and its usage may have impact on a vehicle's safety, \gls*{doip} that is used as transport protocol needs to ensure the diagnostics security.

\subsection{Diagnostic Communication over Internet~Protocol~(DoIP)}\label{sec:doip}
The ISO document 13400~\cite{iso_iso_2011,iso_iso_2019,iso_iso_2016,iso_iso_2016-1} defines the \gls*{doip} protocol as a communication protocol for automotive electronics to transport \gls*{uds} messages.
\gls*{doip} was developed while emissions-related diagnostic data first became mandatory~\cite{iso_iso_2011}.
% Over the years, \gls*{doip} has evolved to handle manufacturer-specific tasks as well~\cite{iso_iso_2011}. 
%
%%TODO: Remove this paragraph?
%Conventional connection architectures such as FlexRay~\cite{autosar_specification_2015} or \gls*{most}~\cite{iso_iso_2020-1} cannot be used to connect directly to the vehicle, so gateways are necessary to convert messages from the in-vehicle networks to a client \gls*{doip} entity~\cite{iso_iso_2011}. Moreover, the Controller Area Network (CAN)~\cite{iso_iso_2015} networks cannot meet the bandwidth requirements that are needed. This is due to their data size of 8 bytes, which results in a maximum bandwidth of 1 Mbps~\cite{ajin_study_2016}. In comparison, Ethernet has bandwidths up to 100 Mbps (100BaseTx Ethernet), which makes diagnostics and software updates possible~\cite{Ajin2016,iso_iso_2011}. For vehicles, there are some requirements for the network system: cost of implementation, available equipment, security, and bandwidth. Ethernet meets most of these requirements~\cite{ajin_study_2016}.
%\medskip
%
The goal of ISO\,13400 is to provide a standardized in-vehicle interface that separates the vehicle network from external test equipment and provides a long-term stable communication interface with the vehicle~\cite{iso_iso_2011}. 
%It should also use industry standards to read diagnostic and manufacturer-specific data and perform tasks for as long as possible~\cite{iso_iso_2011}. It shall be easily adaptable to use new physical and data link layers. 
%To achieve all this, all parts of the \gls*{doip} protocol are built based on the \gls*{osi} basic reference model~\cite{iso_isoiec_1994}. %, which divides communication systems into seven layers~\cite{iso_iso_2011}.
%
To discuss the security of the protocol's design, we establish the basic principles of \gls*{doip}.

\medskip

A \gls*{doip} server typically is a part of an \gls*{ecu} and it responds to requests by a client entity. 
The client typically is an off-board tester but can also be an on-board test device~\cite{iso_iso_2011}. 
%It is possible to have one or multiple clients connected to one vehicle~\cite{iso_iso_2020}.
%
A vehicle can have multiple \gls*{doip} entities and there are several possible network configurations that \gls*{doip} can work in.
The options are (1) to directly connect external test equipment to a vehicle via dedicated physically separate wiring, 
(2) a network connection between one vehicle and one test device,
(3) the connection of multiple vehicles and one tester, 
or (4) one vehicle connected to more than one tester~\cite{iso_iso_2019}.

\medskip

The communication between a client and a server entity using \gls*{doip} can be divided into three phases according to the message payload types:
\begin{itemize}
    \item Vehicle announcement and vehicle identification phase
    \item Routing activation phase
    \item Diagnostic communication
\end{itemize}

\paragraph{Vehicle announcement and identification phase}
Initially, the test device needs to identify each vehicle and its \gls*{doip} entities during the vehicle announcement and identification phase. 
This phase is performed via \gls*{udp} and
can be done in one of two alternative ways regardless of the network configuration~\cite[Figure 11]{iso_iso_2019}:
Either, the \gls*{doip}-server entity, typically a gateway or entity of the vehicle, sends the vehicle announcement message three times as soon as it is connected to the network,
or the test device can request a vehicle identification.
%For instance, this may be the case if the test device missed the vehicle announcement messages because it was not yet ready. 
The tester sends a vehicle identification request message to the vehicle, which responds with the vehicle identification response~\cite{iso_iso_2019}.
The vehicle announcement message and the vehicle identification response contain the same information.
%: \gls*{vin}, \gls*{eid}, \gls*{gid} and the logical address~\cite{iso_iso_2019}.
%
%The group identification (GID) is a decentralized approach to identify \gls*{doip} entities within one vehicle. 
%There is a \gls*{vin}/\gls*{gid} master, from which other entities receive the \gls*{vin}/\gls*{gid} during synchronization process.
One part of the vehicle announcement or identification process is the synchronization feature.
It is intended to synchronize the decentralized \gls*{doip} entities' identification throughout the vehicle.
%The \gls*{doip} entities have preset invalid values before the synchronization is finished.
The exact synchronization method is left to the vehicle manufacturer~\cite{iso_iso_2019}.

\paragraph{Routing activation phase}
After the announcement phase, the tester must establish a \gls*{tcp} connection to the vehicle and then activate the routing.
If a secure connection is requested and supported by the implementation, a \gls*{tls} connection is established.
The \gls*{doip} gateway forwards received data from the diagnostic messages to the \gls*{doip} entities in the vehicle according to the address information~\cite{iso_iso_2019}. 
This is called routing, is done on the vehicle-specific network transport protocol~\cite{iso_iso_2019},
and is unrelated to \gls*{ip} routing that takes place on the \gls*{osi} network layer.

\paragraph{Diagnostic communication}
If the routing activation was successful, diagnostic messages can now be exchanged via the \gls*{tcp} connection. When a \gls*{doip}-server entity receives a message, first, the \gls*{doip} header handler is called. If it is a diagnostic message, it is processed by the diagnostic message handler. 
%The receiver reports the reception of a \gls*{doip} message by a \gls*{doip} confirmation, and the gateway forwards the message to the logically addressed \glspl*{ecu} mentioned in the message. If the received message contains a compliant message, a reply can be sent back. This behavior is specified by the protocol encapsulated in \gls*{doip}~\cite{iso_iso_2019}.
%
The connection termination happens through the \gls*{tcp} teardown mechanism, after which all resources are cleaned up.
If it is not closed regularly after usage, a general inactivity timer or an alive check message can close the connection. 
If a secure connection was used through \gls*{tls}, the socket is closed through the \gls*{tls} mechanisms~\cite{iso_iso_2019}.
%Now the socket is ready for a new connection.

%Beschreibung
With the power mode information request, a client can retrieve if the vehicle is in diagnostic power mode. 
This mode affects functions of the \gls*{doip} servers in the vehicle and its status is reported as either: 
not ready, ready, or not supported.
\textit{Not ready} means that not all servers that can be reached via \gls*{doip} can communicate. 
When the vehicle is in diagnostic power mode, it responds with \textit{ready} and all servers can communicate. 
If the power mode information request message is unsupported, it returns \textit{not supported}. 
The client can use the power mode request message to check if the vehicle is in the correct power mode to perform reliable diagnostics~\cite{iso_iso_2019}.

\begin{figure}
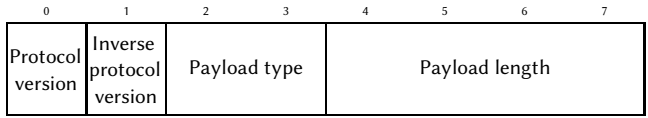
 % [bgcolor=maincolor]
  \sffamily\small
  \begin{bytefield}[bitwidth=0.125\linewidth, bitheight=3.5\baselineskip]{8}
    \bitheader{0-7}\\
    \bitbox{1}{Protocol \newline version} & \bitbox{1}{Inverse \newline protocol \newline version} & \bitbox{2}{Payload type} & \bitbox{4}{Payload length} \\
  \end{bytefield}
  \caption{Overview of a generic DoIP header~\cite{iso_iso_2019}.}
  \label{figure:doip-message}
\end{figure}

\paragraph{Structure of the DoIP messages}
The header of a \gls*{doip} message is located at the beginning of each message and has a total length of 8 bytes. 
\autoref{figure:doip-message} shows the structure of a \gls*{doip} header.
The used protocol version is at the first position and has a length of 1 byte.
At the time of this writing, there are three versions:
\begin{itemize}
  \item \(\mathtt{01}_{16}\): ISO/DIS\,13400--2:2010
  \item \(\mathtt{02}_{16}\): ISO\,13400--2:2012
  \item \(\mathtt{03}_{16}\): ISO\,13400--2:2019~\cite{iso_iso_2019}.
\end{itemize}

Position two contains the inverse protocol number to ensure that a correctly formatted \gls*{doip} message was received.  The next position is the payload type, with a length of 2 bytes, followed by the payload length. The payload length field is limited to 4 bytes, which limits the payload size to 4 GB. 
The maximum allowed payload length is additionally dependent on the transport layer used~\cite{iso_iso_2019}. 

%In the current standard (ISO 13400-2:2019), sixteen different payload types are defined. The document reserves three possible ranges of values for future \gls*{doip} versions and the values \(\mathtt{F000}_{16}\) to \(\mathtt{FFFF}_{16}\) are intended for manufacturer-specific messages. 
%The values of the payload types are grouped threefold: 
%node management (\(\mathtt{XX16}_{16}\)), 
%vehicle information (\(\mathtt{4XXX}_{16}\)), and 
%diagnostic messages (\(\mathtt{8XXX}_{16}\))~\cite{iso_iso_2019}.

\paragraph{Security Requirements}
%The security requirements for which the \gls*{doip} protocol must be examined are defined in this section.
%In a scenario where there is a car and testers, there are several assets that need to be protected: 
%the car's internal network, the tester, and the messages that are sent between the both.
%The tester is critical, since it provides access to the car's internal network and contains various databases with sensitive data.
%\todo{Tester is out of scope.}

\gls*{doip} is used for the transmission of diagnostic data in \gls*{uds} messages and, thus, the security of the diagnostic communication is inherently dependent on \gls*{doip}'s security.
%Especially if \gls*{doip} is used without secure encryption, attacks on the system are conceivable. 
%Among other things, session hijacking or \gls*{mitm} attacks using tampered messages are possible. 
%In addition, \gls*{doip} does not have any intended authentication mechanisms~\cite{matsubayashi_masaru_attacks_2021}.
%
% TODO: Maybe adjust this part more to the AI, autonomous driving theme.
The security properties that are necessary to ensure secure diagnostics are confidentiality, integrity, availability, access control, freshness, and data origin authenticity.
% non-repudiation, 
Confidentiality is essential since an attacker who obtains information and commands by eavesdropping during a diagnostic data transmission can use these to find vulnerabilities and critical information to use in subsequent attacks.
%Integrity implies that the correctness and completeness of data and information is guaranteed over their complete lifetime. 
It is important to ensure integrity so that unauthorized persons cannot manipulate data in the messages and cause unintended behavior that may be harmful.
Interruptions of the availability when the receiver processes and answers \gls*{doip} messages must not endanger any human.
Only legitimate tester devices should be able to read out sensitive diagnostic data to prevent information leakage.
%Non-repudiation is also a critical property: Any action performed by a \gls*{doip} entity must be associated with it.
%Privacy plays a role in \gls*{doip} systems
%in case that diagnostic data contains personal information of the driver or other road users.
The freshness of the messages must be ensured so that no copy of a message can be replayed later in an inappropriate situation.
For example, the same message that is innocuous at the workshop may become dangerous on the road if it is delayed or replayed. 
For data origin authenticity it should be ensured that only verified sources send messages and can also prove that they are verified. 
Otherwise, an attacker could pretend to be an authorized user and send messages that manipulate \gls*{ecu} behavior in an harmful way.

\subsection{Tamarin}
Formal analysis uses software tools to prove the secure design of cryptographic protocols. 
This is done by writing a model of a protocol and its usage. 
Security queries can then be used to check the security properties of the protocol. 
Various prover tools exist for this purpose, such as Tamarin~\cite{basin_tamarin-prover_2022} or ProVerif~\cite{blanchet_automatic_2021}.
As input, they require a model of a security protocol, the properties of the adversary, and lemmas for the required security properties of the protocol~\cite{basin_tamarin-prover_2022}. 
With this information, the prover constructs a proof which includes verification and falsification whether the protocol fulfills the security properties in question. 
Tamarin has an interactive mode for evidence and attack graphs,
where manual and automatic proofs can be started. 
%A complete manual of Tamarin is available in~\cite{basin_tamarin-prover_2022}.

We use Tamarin~\cite{basin_tamarin-prover_2022} to investigate \gls*{doip}, since it is a powerful tool for symbolic modeling and security protocol analysis.
It has been used to formally analyze many protocols, including \gls*{tls}\,1.3~\cite{cremers_automated_2016}. 
In their paper on formal analysis of automotive protocols~\cite{lauser_security_2020}, \citeauthor{lauser_security_2020} analyzed \gls*{secoc} with Tamarin. 
They recommend 
\linebreak[4]
Tamarin for the analysis of further automotive protocols~\cite{lauser_security_2020}.

For \gls*{doip}, \gls*{tls}\,1.2 and \gls*{tls}\,1.3 can be used to provide an encrypted and signed channel, which provides a confidential and authentic data transmission.
\citeauthor{cremers_automated_2016} analyzed \gls*{tls}\,1.3~\cite{cremers_automated_2016} with Tamarin and \citeauthor{houmani_formal_2012} formally analyzed \gls*{tls}\,1.2~\cite{houmani_formal_2012}.
According to their results, \gls*{tls} provides a secure communication channel, ensuring encrypted, authentic, integrity-protected, and fresh\-ness-ensuring data transmission.
The \gls*{tls}-specific key 
\linebreak[4]
exchange can be considered secure~\cite{cremers_automated_2016}. 

\subsection{Attacker Model}

The diagnostic interface via \gls*{uds} is necessary to enable repairs with the large number of \glspl*{ecu} in modern vehicles. 
This is why profound changes to the vehicle can also be carried out via the diagnostic interface. 
Attacks through the diagnostics interface include to
read and delete \glspl*{dtc}, which could lead to greater subsequent damage and thus higher costs for the owner of the car due to the non-detection of a fault. 
It would also be possible to read out sensitive data about the car or even personal data about the owner. 
The deactivation of security-relevant functions is also conceivable. 
For special repairs it is necessary to temporarily switch off safety-relevant functions, such as brakes. 
This is achieved via the car's diagnostic interface in a workshop. 
If an attacker exploits this by preventing the function from being activated after the repair, the vehicle is a danger to drivers and other road users. 
One of the most critical attack vectors is the flashing of unapproved software to the \glspl*{ecu}, which may have different intentions. 
One can attempt to tune the car to get more performance out of it via manipulated software.
This may result in broken safety measures that will result in dangers for the road user.
Thieves could disable important systems like the immobilizer by flashing new software.
Attackers could also deactivate other functions that are necessary for driving. 
All of these diagnostic functions are typically conducted using \gls*{uds}.

For such attacks, the adversary requires to modify or in some cases only read from the network communication.
Such an attacker may be eavesdropping or performing session hijacking during a running diagnostic session in vehicle production or in a workshop, which may violate confidentiality and integrity.
During normal operations, i.\,e., the regular car usage when driving on the road, attacks may include replay or spoofing of valid messages using the attackers capability to inject messages into the network, which primarily violates the integrity with potential safety impact.
This adversary is approximated by the Dolev-Yao attacker.
They can read, modify, intercept, and inject new messages on the network.
%For a stronger adversary, custom multiset rewriting rules must be written. 
%For example, to give the attacker the ability to reveal long-term-keys, the user can model this with a rule by sending the long-term key out into insecure network. 
The standard attacker in Tamarin models is a Dolev-Yao adversary, which we use in our models in \autoref{sec:formal-analysis}.

\section{Security Analysis}\label{sec:security-analysis}
% !TeX spellcheck = en_US
We organize the deductive security analysis in this section and the formal analysis using Tamarin in the subsequent section into subsections that resemble the \gls*{doip} protocol phases: 
the vehicle announcement and identification phase, 
the routing activation, and 
the diagnostic communication. 
We prepend a generic consideration of the security of the \gls*{doip} header to the discussion of the phases.
We analyze the specification of ISO\,13400--2~(2019)~\cite{iso_iso_2019} and compare it to the results of the analysis~\cite{lindberg_security_2011} of the draft version from 2011.
Our analysis is the first to consider the addition of \gls*{tls} to the standard, which is the major change between the standard versions.
All references to the standard are based on the latest version from 2019.
In this section, we refer to tables, state machines, as well as requirements in the standard using the naming scheme \textit{DoIP-xxx} that is used in the ISO\,13400--2 document.

\subsection{DoIP Header}\label{sec:doip-header}

\subsubsection{Existing Measures}\label{sec:header-existing-measures}
When a \gls*{doip} entity receives a message, it first calls the \gls*{doip} header handler~\cite[Figure 16]{iso_iso_2019}. 
Some mechanisms are built into the \gls*{doip} header handler to protect the \gls*{doip} entities. 
According to requirement DoIP-031~\cite{iso_iso_2019}, each \gls*{doip} entity shall ignore packets that have a multi- or broadcast address as the source \gls*{ip} address. 
This helps to prevent an amplification attack: 
One malicious broadcast packet would trigger one reply packet from \emph{each} node in the network.

The requirements DoIP-039~\cite{iso_iso_2019} and DoIP-040~\cite{iso_iso_2019} are intended to prevent NACK storms: 
\gls*{doip}-client entities shall not respond to an invalid \gls*{doip} message from a \gls*{doip} entity with a NACK in the header.
Sending NACK messages is only allowed for the \gls*{doip}-server entities, but not for the clients.
Servers ignore incoming negative acknowledged messages.
The standard specifies exactly when a message should be discarded, for example when a message is too large and a buffer overflow would be possible.
This measure prevents overloading the network and server capacities.

\subsubsection{Vulnerabilities}\label{sec:header-vulnerabilities}

The header contains the protocol number and its inverse, to ensure that correctly formatted messages are received. However, the two protocol numbers are only 2 of the 8 bytes of the header. Therefore, it is a poor integrity check, not only from a security perspective. According to \citeauthor{lindberg_security_2011}, this was already the case in the draft version and has not changed with the current version~\cite{lindberg_security_2011}.

\subsection{Vehicle Announcement and Vehicle~Identification}

\subsubsection{Existing Measures}\label{sec:vi-existing-measures}
The specification limits the number of vehicle announcement messages sent into the network to three.
By limiting the number of repeated transmissions, a denial of service vulnerability by toggling the connections of \gls*{doip} entities on and off is mitigated if an attacker has not enough control over the entities to unrestrictedly send messages into the network themselves. 
This is a weak mitigation for a weak kind of attacker that has severely limited control over the network and any of its entities.

As specified in DoIP-051~\cite{iso_iso_2019}, the \gls*{doip} entities send the vehicle identification responses after a random waiting time. 
Otherwise, it would be possible to overload the network, since all \gls*{doip} entities would respond to a vehicle identification request at the same time.
This is an effective functional network protection mechanism rather than a security measure.

\subsubsection{Vulnerabilities}\label{sec:vi-vulnerabilities}
In the vehicle identification phase, no authentication is provided, which is why it is possible for attackers to send false information as a response. 
The adversary does not even have to establish a \gls*{tcp} connection for this, since this phase runs via \gls*{udp}. 
They can respond to requests with their own \gls*{ip} address and pretend to be a vehicle. 
Also lacking any integrity protection, all fields in the response can be changed by an adversary. 
This vulnerability was already present in the draft version~\cite{lindberg_security_2011}.

Figure~10 and Section~6.3.2 in the \gls*{doip} standard~\cite{iso_iso_2019} differ
%\linebreak[4]
in the description of the procedure when a \gls*{doip} entity has the sync status \emph{incomplete}. 
According to the standard's Figure~10~\cite{iso_iso_2019}, vehicle 
\linebreak[4]
identification request messages are sent again after the 
\linebreak[4]
\texttt{vehicle\_discovery\_timer} has expired. 
In the textual description, a request should only be sent to the entity that has set the sync status to \emph{incomplete}. 
Such an inconsistency in the definition may allow adversaries to fingerprint which implementation is running on a system. 
If an implementation has a particular vulnerability, this can targetedly be exploited~\cite{lindberg_security_2011,iso_iso_2019}. 
Besides this still-present inconsistency, further inconsistencies between individual definitions existed in the draft version~\cite{lindberg_security_2011}, but these are no longer contained in the current version.

Another vulnerability is associated with the sync status: 
it can be exploited for a denial-of-service attack. 
The attacker is an additional node in the network and sends spoofed vehicle identification responses with sync status set to \emph{incomplete}. 
Because of this message, the client needs to wait for the server to be synced and then start again with a vehicle identification request.
As client and servers are affected  by this attack, the vehicle identification fails
and the tester cannot connect to the vehicle. 
This attack~\cite{lindberg_security_2011} was already possible in the draft version.

\subsection{Routing Activation}

\subsubsection{Existing Measures}\label{sec:ra-existing-measures}
After the vehicle identification phase and the establishment of a \gls*{tcp} connection, the client must activate routing.
At this point, \gls*{tls} and thus encryption can be used for the first time. 
As long as routing is not activated, the \gls*{doip} server ignores all other messages except the routing activation request message. 
Also unknown \acrlongpl*{sa} are not accepted and the socket is closed again. 
While slightly increasing an attackers effort, it is easy so spoof \acrlongpl*{sa} to circumvent this measure.

Optionally, for better protection, the car may require extra measures to accept routing activation from a tester. 
These include manual user consent, confirming the routing activation in the car cockpit screen to prevent unwanted connections. 
In this case, to proceed the attacker needs access to the car's interior~\cite[DoIP-105]{iso_iso_2019}. 
\gls*{tls} can also be required by the car for certain diagnostic
\linebreak[4]
sessions~\cite[DoIP-174]{iso_iso_2019}.
Manual user consent and \gls*{tls}, both, are very effective measures for mitigating numerous vulnerabilities, which considerably increase the complexity of an attempted attack.
\autoref{tab:vns} gives an overview of which vulnerabilities are mitigated by the use of \gls*{tls}.

Figure~22 in the standard~\cite{iso_iso_2019} mentions an optional authentication mechanism in the \gls*{doip} protocol during the routing activation phase. 
This authentication mechanism is not specified in detail. 
The realization is left to the vehicle manufacturers and the standard recommends higher layers for it. 
Due to the lack of a precise specification, we cannot discuss the effectiveness of the optional authentication.
If adequate authentication is required from the client, an attacker cannot activate the routing.

%\subsubsection{Socket handler}
Socket handling is part of the routing activation handler. It 
\linebreak[4]
checks how many sockets are currently connected and if the \acrlong*{sa} is already used by another socket. 
If there is still a socket available and the same \acrlong*{sa} is not yet registered with another socket, it is registered with the socket.
If a new connection is established, a new initial inactivity timer is started for the corresponding \acrlong*{sa}. 
When the client does not send a routing activation request message while the timer is running, the socket is closed again. 
If the client sends a routing activation request message in time, the initial inactivity timer is stopped and the general inactivity timer is started. 
These two timers provide protection against resource exhaustion attacks. 
An attacker who occupies sockets must do this actively, otherwise the inactivity timers close the socket again. 
A single \acrlong*{sa} can only connect to one socket, so an attacker can only occupy one socket with one \acrlong*{sa}.
However, \acrlongpl*{sa} are easy to spoof for circumventing this measure.

\subsubsection{Vulnerabilities}\label{sec:ra-vulnerabilities}

\Acrlongpl*{sa} are checked for validity 
\linebreak[4]
prior to the connection setup during routing activation.
If the provided \acrlong*{sa} is unknown, this information is returned in the response. % with the code \(\mathtt{00}_{16}\). 
An attacker is therefore able to scan for valid \acrlongpl*{sa} to gain knowledge about clients known to the vehicle.
This check is performed before the aforementioned optional authentication and before the connection setup, 
so even with activated authentication and \gls*{tls} this information leak is not not mitigated. 
%Client authentication with \gls*{tls} is optional in \gls*{doip} and according to the standard, authentication of the client should be performed at a higher layer, for example in \gls*{uds}~\cite[Section 10]{iso_iso_2019}. 
This kind of scanning for \acrlongpl*{sa} was already possible with the draft version~\cite{lindberg_security_2011}.
This vulnerability may be of low criticality since the only information the attacker gains is the address of the tester that currently has requested to connect to the vehicle diagnostic interface at this specific point in time.

The \gls*{doip} server can reject the routing if \gls*{tls} is necessary. 
This is documented in the requirement DoIP-174~\cite{iso_iso_2019}. 
However, for the routing activation handler this case is not shown in Figure~22 of the DoIP standard~\cite{iso_iso_2019}. 
Similar to the vulnerabilities in the vehicle identification phase, this deviation of specifications in the standard can lead to fingerprinting. 
It is not stated when it should be checked in the handler whether a \gls*{tls} connection is necessary. 
This inconsistency is new in the current DoIP version.

If the vehicle requires authorization during the routing activation phase but \gls*{tls} is not used, an adversary can hijack the \gls*{tcp} connection after the routing activation phase and thus bypass authorization. 
However, this requires a previously active connection to the vehicle to hijack.

\subsection{Diagnostic Communication}
The diagnostic communication is the phase after the server activated the routing for the client. 
The four different message types of this phase are the 
\begin{itemize}
\item diagnostic messages, 
\item alive check messages, 
\item power mode information messages, and 
\item \gls*{doip} entity status information messages.
\end{itemize}
%
% Vuln.
There is no authentication in the diagnostic message handler and, without an encrypted and integrity protected channel, 
an adversary could try to intercept and modify messages to inject potentially dangerous payload into the vehicle.

\subsubsection{Diagnostic messages}\label{sec:diagnostic-messages}
% CM
The diagnostic message is a message type for routing diagnostic requests to and responses from the vehicle network. 
Diagnostic messages are confirmed with a positive or negative ACK. 
As explained in \autoref{sec:header-existing-measures}, only messages sent by the client are confirmed to prevent NACK storms.

% TODO interesting but needs more details and is not about "the protocol" itself
%The diagnostic message handler has built-in checks to prevent buffer overflows~\cite[Figure 17]{iso_iso_2019}.
%\medskip

Error messages sent from the \gls*{doip} server entity can be used to scan the target addresses in the vehicle. 
The client can request target addresses and receives an error message if it is not known to the server. 
The attacker thus gains information about the in-vehicle network. 
These error messages already existed in the draft version~\cite{lindberg_security_2011}.

\subsubsection{Alive check message}\label{sec:alive-check-message}
The alive check message is used to ascertain if a \gls*{tcp} socket is still in use. 
%The socket handler uses this message type.
Two types of vulnerabilities exist in this mechanism:

% Vuln.
To exploit the first, the adversary needs to intercept the alive check request messages of the server or the alive check response message of the client. 
Either way, the server receives no response to its request, thinks the socket is no longer active, and closes it.

Secondly, an attacker has the possibility to send alive check response messages when the client has already disconnected from the server. 
The server still thinks the socket is in active use and the socket stays open and remains unavailable for new connections. 
However, this requires to hijack the \gls*{tcp} session beforehand to keep the channel open and to prevent the \gls*{tcp} teardown. 
With \gls*{tls}, both types of attacks are no longer possible, since interception of specific messages and hijacking is not feasible if they are encrypted. 
The attacks on the alive check messages were already possible in the draft version~\cite{lindberg_security_2011}.

\subsubsection{Power mode information}\label{sec:power-mode-information}
% vuln.
The possibility of an unnoticed manipulation of the vehicle's power mode information response causes the following vulnerability. 
A denial of service takes place if a test device sends a power mode information request to the connected vehicle and receives a spoofed power mode information response. 
If the test device is mislead that the vehicle is not ready, the tester does not proceed.
In contrast, if the test device is made to believe that the car is ready, although at least one server is not yet ready, diagnostic of the car may fail.
Besides these denials of service, a second kind of vulnerability may arise in the latter case.
Since, from a functional point of view, it should not be possible that critical diagnostic requests take place if \glspl*{ecu} are not ready, this may cause an unspecified operation state.
If this is not handled graciously, the implementation may react in an undesirable way to this non-well specified state the attacker places the protocol in.
An unjustified reaction in a normal operation mode may cause even safety impact if no additional precautions to deal with this undefined situation are taken.

With \gls*{tls} the power mode information's integrity is protected and the attack is no longer possible. 
Again, this was also possible in the draft version~\cite{lindberg_security_2011}.

\subsubsection{DoIP entity status information}\label{sec:doip-entity-status-information-message}
%Beschreibung
The \gls*{doip} entity status information message is used to retrieve information about the other \gls*{doip} entity in one connection. 
The response contains information like the entity type, maximum number of concurrent \gls*{tcp} sockets and how many of them are currently open, as well as the maximum data size. 
The entity type indicates whether it is a \gls*{doip} node or a \gls*{doip} gateway.
The following vulnerability of the status information were already present in the draft version~\cite{lindberg_security_2011} and still are in the current standard. 

% vuln.
Message spoofing is possible with all four fields of the response message with the intention to disrupt or prevent the exchange of diagnostic data.
Changing the entity type or
the information about the sockets 
may prevent diagnostics from running correctly or at all.
If the adversary replaces the maximum data size in the response by a larger number, this can result in the diagnostic message handler of the server discarding the message because it is too large. 
Protected by \gls*{tls}, this is impossible and values of the messages cannot be manipulated by an adversary.

\subsection{Evaluation of the Results}

\begin{table*}[]
    \caption{Vulnerabilities of the \gls*{doip} protocol.}
    \label{tab:vns}
    \small
    \begin{tabularx}{\textwidth}{Xllllcr}
        \toprule
        \textbf{Vulnerability}                                   & \textbf{Violates} & \textbf{Target}      & \textbf{Type}       & \makecell[l]{\textbf{Attacker}\\ \textbf{capability}} & \textbf{Section}                            & \makecell[r]{\textbf{Mitigated}\\ \textbf{by \gls*{tls}}}          \\
        \midrule
        Weak protection of version number integrity            & integrity         & client + server      & downgrade           & inject packet           & \ref{sec:header-vulnerabilities}                 & partly \\ % \makecell[l]{partly (diagnostic\\ communication)}
        Missing authentication in vehicle identification & authorization     & client               & illegitimate access & inject packet           & \ref{sec:vi-vulnerabilities}                     & no                                       \\
        Inconsistency in sync status                             & confidentiality   & server               & fingerprinting      & read from network       & \ref{sec:vi-vulnerabilities}                     & no                                       \\
        Spoof failed sync status                                 & availability      & client               & denial of service   & inject packet           & \ref{sec:vi-vulnerabilities}                     & no                                       \\
        Scanning for \acrlongpl*{sa}                              & confidentiality   & server               & information leak    & inject packet           & \ref{sec:ra-vulnerabilities}                     & no                                       \\
        Inconsistency in \gls*{tls} activation                    & confidentiality   & server               & fingerprinting      & read from network       & \ref{sec:ra-vulnerabilities}                     & no                                       \\
        \gls*{tcp} hijacking during routing   activation          & authorization     & \gls*{tcp} connection & illegitimate access & inject packet           & \ref{sec:ra-vulnerabilities}                     & yes                                      \\
        Scanning for target addresses                            & confidentiality   & server               & information leak    & inject packet           & \ref{sec:diagnostic-messages}                    & yes                                      \\
        Intercept/inject alive check messages                    & availability      & server               & denial of service   & inject packet           & \ref{sec:alive-check-message}                    & yes                                      \\
        Spoofing of power mode information                       & integrity         & server               & illegitimate access & inject packet           & \ref{sec:power-mode-information}                 & yes                                      \\
        Spoofing of status information                   & availability      & server               & denial of service   & inject packet           & \ref{sec:doip-entity-status-information-message} & yes                                      \\
        \bottomrule
    \end{tabularx}
\end{table*}

The vulnerabilities that exist in \gls*{doip}, which \autoref{tab:vns} summarizes, can be divided into three main types:
Vulnerabilities that (1) leak information about the vehicle network and its \gls*{doip} entities, 
such that lead to (2) gaining illegitimate access, and (3) denials of service.
All attacks exploiting vulnerabilities in the \gls*{doip} header, the vehicle announcement and vehicle identification phase, and routing activation phase are still possible even with an activated \gls*{tls} or the optional authentication during routing activation.
Attacks on existing diagnostic connections are as diverse and critical as those during the vehicle identification and routing activation phases.
However, these attacks can be mitigated by using a secure connection provided by \gls*{tls}, which prevents interception and manipulation of messages.
Activating the optional \gls*{tls} is an effective countermeasure, since for these attacks the attacker needs to inject valid packets into the channel.
% Mitigation by TLS
However, this mitigation is only sufficient if authentication is used as early as during routing activation and allows only authorized \gls*{doip} entities to participate in the communication. 
Thus, \gls*{tls} and authentication should be enforced for every diagnostic session.
However, this may cause incompatibilities with older versions of the \gls*{doip} protocol that did not support \gls*{tls} and authentication.

Information about the \gls*{doip} implementation used by a specific entity can be gained from inconsistencies in the standard's descriptions.
These inconsistencies allow fingerprinting, i.\,e., an attacker can identify the implementation used, in preparation of an imple\-mentation-specific attack.
The specification in the standard has been improved since the 2011 draft version, but two major inconsistencies still exist in the current version:
One in the sync status specification and one in the \gls*{tls} activation specification.

\medskip

We discuss the fulfillment of the security requirements of \gls*{doip}, described in \autoref{sec:doip}.
While \gls*{tls} is now supported by \gls*{doip}, its use is not mandatory and \gls*{doip} can still be used without any protection. 
Thus, we discern the security properties of \gls*{doip} with and without \gls*{tls}.

The \gls*{doip} protocol specifies to use only the \acrlongpl*{sa} to determine the origin of data.
This is insufficient to provide \emph{authenticity} and \emph{authorized access}, since spoofing and connection hijacking vulnerabilities exist for the insecure connection.
%This also thwarts non-repudiation.
Without a secure connection, for the lack of any kind of message verification mechanism like signatures, the \emph{integrity} of the diagnostic data is also unprotected.
Likewise, no mechanism in \gls*{doip} can determine whether a message is current to prevent replay attacks.
\emph{Freshness} is therefore not guaranteed without a secure connection. 
Without encryption, the \emph{confidentiality} in the system is not ensured.
The \emph{availability} is not guaranteed, since denial of service attacks that exploit the lack of integrity protection are possible.

With \gls*{tls} in use for providing a secure channel,
\emph{confidentiality}, \emph{integrity}, \emph{authenticity}, and \emph{freshness} are gained by the secured channel during the diagnostic communication.
The lack of a possibility to use \gls*{tls} in the earlier protocol phases prevents any improvement of the protection of the communication during these phases.
If the client connects via an otherwise secure channel, the authorization in the routing activation phase is sufficient to guarantee \emph{authorized access} to the vehicle network. 
\emph{Authorization} can be gained through mutual authentication of the client and server during \gls*{tls} connection setup in the routing activation phase.
The fine-grained access control to \gls*{uds} diagnostics must be handled in the application layer.

\medskip

By its design, no dedicated security measures are built into the \gls*{doip} diagnostic communication phase.
\gls*{tls} was added only in the latest version of the standard and was not part of the original protocol design.
The design of the vehicle announcement, vehicle identification, and parts of the routing activation phases does not even allow the use of \gls*{tls}.
Furthermore, the existing measures in the vehicle announcement, vehicle identification, and routing activation phases do not sufficiently mitigate the vulnerabilities, specifically in the absence of the possibility to communicate over any secure channel in these protocol phases.
An attacker that can be mitigated by the existing measures, which we described in \autoref{sec:header-existing-measures} and \autoref{sec:vi-existing-measures}, has only very limited access to the network and cannot inject arbitrary packets into the network, as a more realistic Dolev-Yao attacker can.
Measures with limited effectiveness against denial of service exist for the vehicle announcement and routing activation phases.
Only the measures \emph{manual user consent} and \emph{authentication}, which both are marked optional in the standard, render sufficient protection against illegitimate access.

\section{Formal Analysis}\label{sec:formal-analysis}
% !TeX spellcheck = en_US
In the preceding security analysis, we found vulnerabilities like information leakage and illegitimate access in the protocol, which we confirm in a formal analysis.
Errors in the design, such as inconsistent specification in the protocol, cannot be detected using formal analysis since 
this information cannot be formalized in the prover model.

The Tamarin models are simple, since there are hardly any security measures in the \gls*{doip} protocol~\cite{iso_iso_2019}, which can be verified. 
We check the models for authenticity and secrecy and if they are fulfilled, we conclude that the integrity and confidentiality properties are fulfilled.
Preventing the information leakage and illegitimate access vulnerabilities, 
a secret message is noted in the model with the action fact \texttt{Secret} as shown in \autoref{lst:secret}.
If there is no case in which the adversary \texttt{K} knows the message, then the message is confidential.
\begin{lstlisting}[caption={Tamarin lemma for the secrecy of a message~\cite{basin_tamarin-prover_2022}.}, label=lst:secret, float=b]
    lemma secrecy:
        "All x #i. Secret(x) @i 
        ==> not (Ex #j. K(x)@j)"
\end{lstlisting}
The Tamarin action fact \texttt{Authentic} is used to verify that the message was not sent by the adversary.
For each message sent, a \texttt{Send} action fact must be modeled as shown in \autoref{lst:authentic} to check the authenticity.
\begin{lstlisting}[caption={Tamarin lemma: authenticity of a message~\cite{basin_tamarin-prover_2022}.}, label=lst:authentic, float=b]
    lemma message_authenticity:
        "All b m #i. Authentic(b,m) @i
        ==> (Ex #j. Send(b,m) @j & j<i)"
\end{lstlisting}

Availability is not investigated in the models.
However, results of the lemma for authenticity allow conclusions about this property.
We cannot verify the access control and freshness properties in the Tamarin models because no respective mechanisms are sufficiently defined in \gls*{doip}.
The Dolev-Yao adversary's control over the network is modeled with two built-in \texttt{In} and \texttt{Out} rules in Tamarin.
% In und Out Facts
% From https://tamarin-prover.github.io/manual/book/005_protocol-specification-rules.html
% In: This fact is used to model a party receiving a message from the untrusted network that is controlled by a Dolev-Yao adversary, and can only occur on the left-hand side of a rewrite rule.
% Out: This fact is used to model a party sending a message to the untrusted network that is controlled by a Dolev-Yao adversary, and can only occur on the right-hand side of a rewrite rule.

We compare the usage of \gls*{doip} with and without \gls*{tls}
but do not verify \gls*{tls} itself in this work, since it has already formally been analyzed~\cite{cremers_automated_2016,houmani_formal_2012}.
The \gls*{tls}-specific key exchange is not represented in the Tamarin models, since \citet{cremers_automated_2016} already formally analyzed it with Tamarin. 
Based on this, we assume for the Tamarin models that \gls*{tls} provides a perfectly secure communication channel.
In case \gls*{tls} is used, we add an authenticated and encrypted communication channel for \gls*{doip} in the analysis where the standard specifies this channel.
We model the secure channel in Tamarin using multiset rewriting rules. 
In total, there are two rules to convert outgoing messages into incoming messages. 
The two rules tightly link the sender and the receiver to the message~\cite{basin_tamarin-prover_2022}. 
By the properties of the secure channel, neither can an adversary manipulate the \gls*{doip} message, nor can they send messages into the secure channel. 
We model to send a message from \texttt{A} to \texttt{B} by the fact \texttt{Out\_S(\$A, \$B, message)}. 
The fact \texttt{In\_S(\$A, \$B, message)} represents \texttt{B}'s reception of the message. 
Consequently, the secure channel of \gls*{tls} is confidential and authentic~\cite{basin_tamarin-prover_2022}. 
In our analysis, we do not examine the authentication and freshness that \gls*{tls} provides.

\subsection{Vehicle Announcement and Vehicle~Identification}
In the vehicle announcement phase there are no formalizable security measures, no encryption or signing, present.
Our Tamarin models show that messages are not secret, nor are the messages authentic and 
all vehicle announcement messages are sent as \gls*{udp} messages to the broadcast address. 
Everybody, including the adversary in Tamarin, can read them. 
The Dolev-Yao adversary has full control over the network and can send messages themselves, pretending to be a \gls*{doip} entity. 
The authenticity of the messages cannot be assured.

A generic vehicle identification request requires a vehicle to respond to this message with a vehicle identification response. 
Since this is also sent unencrypted via \gls*{udp} into the network, anyone can read and modify it, which our Tamarin model recognizes by the failing of the secrecy lemma: 
Tamarin finds counterexamples. 
This confirms our assessment from \autoref{sec:vi-vulnerabilities}, containing multiple vulnerabilities.

\subsection{Routing Activation phase}
A \gls*{tcp} connection is established for routing activation. 
In this phase, \gls*{tls} can optionally be used for the first time. 

Without \gls*{tls} there is neither encryption nor authentication, and thus no security measures, in effect and 
the Tamarin lemmas for message authenticity as well as message secrecy fail. 
As discussed in \autoref{sec:ra-vulnerabilities}, this leads to information leakage and illegitimate access vulnerabilities.

With \gls*{tls} in place, the lemmas on secrecy and authenticity of the messages succeed. 
Spoofing or eavesdropping of \gls*{doip} messages is no longer possible with a secure connection over \gls*{tls}.
This also means that hijacking of connections and bypassing of the authentication in the car is no longer possible. 
\gls*{tls} successfully provides confidentiality and integrity for \gls*{doip} in this phase. 
Still, scanning of \acrlongpl*{sa} is possible with \gls*{tls} in use.

\subsection{Diagnostic Communication}
When routing is activated, client and server can exchange further messages. 
A \gls*{tcp} connection, optionally with or without \gls*{tls} between the tester and the server, can be set up.
There is no authentication specified in this protocol phase. 
The multiple models and message types of this phase are similar to each other.

\subsubsection{Diagnostic messages}
A client that sends a diagnostic message to the server receives an ACK or a NACK from the server. 
When transmitting diagnostic data without \gls*{tls}, there are no security measures that can formally be represented in the model. 
The lemmas for secrecy and message authenticity fail. 
Spoofing and eavesdropping of data is possible. 
This also affects diagnostic data of the higher layers, as long as they do not encrypt their payloads themselves.

If \gls*{tls} is used, the lemmas succeed, so spoofing of and eavesdropping on messages is impossible. 
The authentication provided by \gls*{tls} in the routing activation phase is sufficient to achieve the authentication security property.

\subsubsection{Alive check messages}
The situation is similar for alive check requests and responses. 
There are no verifiable security measures except for the optional \gls*{tls}.
If it is not used, the lemmas fail here as well and spoofing and eavesdropping are possible
and intercepting messages to produce a subtle denial of service attack is possible.

The use of \gls*{tls}, and thus the transmission of data in a confidential, integrity-protected, and authenticated channel, prevents eavesdropping, unnoticed interception, and modification of messages. 
The lemmas for secrecy and authenticity of messages succeed. 
Adversaries can no longer perform denial of service attacks through the alive check messages. 
\gls*{tls} provides the means to ensure the availability property in case of the spoofing vulnerability in this phase described in \autoref{sec:alive-check-message}.

\subsubsection{Power mode and entity status information}
The power mode information and the entity status information also lack any security measures.
The lemmas for secrecy and authenticity fail. 

With \gls*{tls}, the lemmas succeed, showing that it can fix the spoofing vulnerability, which is the only existing one for these status messages.

\subsection{Summary}
The formal analysis confirms the results of the security analysis of the \gls*{doip} protocol in \autoref{sec:security-analysis}. 
Due to the lack of any formally verifiable security measures, eavesdropping and manipulation of messages is possible without the use of \gls*{tls}. 

In contrast, the lemmas for secrecy and authenticity of messages do not fail with \gls*{tls} in place.
Thus, the Tamarin models show that \gls*{tls} provides ample security for a protocol that contains no security measures as of itself.
Thus, it should be deprecated to omit \gls*{tls} for \gls*{doip}.
We discuss further countermeasures in the next section.
\bigskip
\bigskip

\section{Recommended Countermeasures}
% !TeX spellcheck = en_US
This chapter presents our proposal for measures to improve the security of the \gls*{doip} protocol.
The measures are based on the results from \autoref{sec:security-analysis} and the results of the formal analysis in \autoref{sec:formal-analysis}.

\medskip

To add any confidentiality, authenticity, and authentication, the use of \gls*{tls} should be mandatory for \gls*{doip} connections over a network. 
The standard currently allows that the vehicle manufacturer decides if a secure connection is necessary for only certain diagnostic sessions. 
To improve security for all connections, 
% except a dedicated connection between tester and vehicle (but then still: a injected device), Thus: remove this sentence 
a secure connection should be chosen by the manufacturer. 
If used, \gls*{tls} secures the messages between client \gls*{doip} entity and server \gls*{doip} entity~\cite{iso_iso_2019}.
According to the standard, \gls*{doip} entities should support \gls*{tls}\,1.2 and 1.3.
However, only \gls*{tls}\,1.3 provides a state-of-the-art security level. 
For example, authentication-only ciphers are accepted in \gls*{doip} with \gls*{tls}\,1.2.
The standard states that these should exclusively be used for debugging and development and should be deactivated for operational communication. 

The negotiation of a cipher suite between the entities follows the \gls*{tls} specification:
The client starts with asking for the highest \gls*{tls} version it supports. 
The server then responds with the same version or has the option of choosing a different version if it does not support the requested one~\cite{iso_iso_2019}. 
Since attacks are known on \gls*{tls}\,1.2 and its cipher suites~\cite{sheffer_summarizing_2015},
the vulnerable ones should be removed from the list of supported cipher suites for \gls*{doip}.
Even better would be to only rely on \gls*{tls}\,1.3. 

Using \gls*{tls} prevents spoofing and eavesdropping of messages.
Denial of service attacks based on manipulated or intercepted messages can also be prevented. 
The use of \gls*{tls} prevents attacks on the routing activation phase and all subsequent messages during diagnostic communication.

\subsection{DoIP Header}\label{sec:cm-doip-header}
In the \gls*{doip} header, the inverse protocol version is used as an integrity check to verify that correctly formatted messages are received. 
Instead of using the inverse protocol number, the standard should use a signature over the complete header. 
This would then cover all bytes of the header, and not only the first two,
which we consider a weak kind of check. 
A signature provides a simple and reliable way to verify that a correct message was received unaltered.

\subsection{Vehicle Announcement and Vehicle~Identification}
The vehicle announcement phase is the most difficult to realize countermeasures for. 
\gls*{udp} is used and messages are sent to broadcast addresses, which is why \gls*{dtls} is not an option: 
While \gls*{dtls} creates a secure channel for \gls*{udp}, it does not support broadcasts.
A signature as proposed in \autoref{sec:cm-doip-header} would mitigate the synchronization status spoofing and thus is our recommendation.
In addition, the ambiguities in the specification need to be fixed.

\subsection{Routing Activation Handler}
Authentication can be used in the routing activation phase. 
This authentication is optional and its realization is left to the vehicle manufacturer by the current ISO standard. 
The standard recommends that authentication should be performed on the application layer.
Instead, we argue that an optimal approach would be to use \gls*{tls} client authentication. 
Together with \gls*{tls}, it would significantly improve the security of the system, 
since it reduces the overall complexity:
A confidential, integrity-protected, and mutually authenticated channel is established from the beginning 
and authorization in the application is based directly on the authentication of \gls*{tls} without a separate mechanism or additional credentials.

To prevent \acrlong*{sa} scanning, the sequence of the routing activation handler should be changed. 
The routing activation handler should first authenticate the client, and only after a successful authentication check whether the source address is known. 
This would prevent scanning of \acrlongpl*{sa} and at the same time also fingerprinting of the implementation by an unauthenticated attacker.

\subsection{Diagnostic Messages}

By using \gls*{tls} and \gls*{tls} client authentication, adversaries cannot connect without a valid client certificate and thus cannot enumerate target addresses.
As we argue above, the subsequent authorization for diagnostic services would be most elegant if using the already performed \gls*{tls} client authentication. 
Spoofing and eavesdropping of messages after the routing activation phase can only be prevented by using a \gls*{tls} connection. 
With these two measures, all vulnerabilities in the diagnostic communication phase can be mitigated and all desired security properties in this phase can be achieved.

\subsection{Summary}
%
%\begin{table*}
%    \begin{tabularx}{\textwidth}{llX}
%        \toprule
%        Security property & Status & Description                                     \\
%        \midrule
%        
%        Data origin authenticity & Fulfilled & Fulfilled when using \gls*{tls}.      \\
%        Integrity                & Fulfilled & \gls*{tls} provides integrity.        \\
%        Authorization            & Fulfilled & Fulfilled when using \gls*{tls} client authentication. Can be supported by authentication on the application layer. \\
%        Freshness                & Fulfilled & Replay attacks are not possible with \gls*{tls}. \\
%        Non repudiation          & Fulfilled & Origin of messages is known with \gls*{tls}.     \\
%        Confidentiality          & Fulfilled & Fulfilled by \gls*{tls}.                         \\
%        Availability             & Broken    & Still broken because of the vehicle identification phase. But the other existing attacks are no longer possible with the countermeasures.\\
%        \bottomrule
%    \end{tabularx}
%    \caption{Security properties after adding the countermeasures. \todo[inline]{compatibility of CM, before/after CM/TLS}}
%    \label{table:countermeasures-summary}
%\end{table*}

The proposed measures could significantly improve the security of \gls*{doip}. 
Data origin \textbf{authenticity} is satisfied by using \emph{\gls*{tls}} and \emph{client authentication}, so 
only verified sources can send messages and establish a connection. 
Still vehicle identification requests and responses can be sent by unauthenticated entities. 
If we assume an access control is based on the \emph{\gls*{tls} client authentication}, the security property \textbf{authorization} is also fulfilled.

While dumb attacks on the availability are always possible on a system under the attacker's physical control, like destroying, unplugging, or jamming, more subtle attacks on the availability can be prevented.
\textbf{Availability} can be improved by an \emph{authenticity check} for the identification phase.
Thus, an adversary could not forge a synchronization message to prevent the entities from connecting.
To implement this, the \gls*{doip} protocol would have to be modified, creating an incompatibility between versions.

A \gls*{pki} is required for \gls*{tls} client authentication and the signatures in the \gls*{doip} header of the vehicle announcement, vehicle identification, and routing activation messages that we propose.
Since the effort of managing a \gls*{pki} is already required to activate \gls*{tls} even with only server certificates, the overhead is reasonable.
Access levels that distinguish a common workshop tester with uncritical rights from a development or debugging tester with enhanced rights are part of the service~\(\mathtt{29}_{16}\) concept in ISO\,14229~\cite{iso_iso_2020}.
There, the \gls*{uds} standard describes how a tester can authenticate on application layer towards the vehicle based on a client certificate.
We expect that any infrastructure for certificate management including online tester-revocation mechanisms, protected key storage in tamper-resistant hardware, and secure deployment of client certificates for \gls*{uds} can be used for \gls*{doip} instead.
Beyond currently existing standard solutions, in the future \gls*{uds} could then be provided with a delegate of the \gls*{doip} certificate to implicitly be authorized on the application layer.

\section{Conclusion}
% !TeX spellcheck = en_US
%% TODO: Remove all the medskips and other skips.

We analyzed to which extent the DoIP protocol is a secure transport protocol for diagnostic data.
Therefore, we performed a comprehensive security assessment, verified by the formal analysis of the security properties of the protocol using Tamarin.

Despite its crucial role in ensuring a secure connection between testers and vehicles,
\gls*{doip} as of itself is insecure in its current state and vulnerabilities exist.
Besides \gls*{tls}, the currently specified measures against denial of service and spoofing in the standard significantly mitigate attacks of adversaries with only limited access to the network. % specificly with a limited sending rate
Because of this, our formal analysis with Tamarin can mainly confirm rather obvious assessments of the security of \gls*{doip}.
Even if the application layer over \gls*{doip} uses its own encryption during diagnostic communication, information leakage from the \gls*{doip} protocol as well as denial of service attacks are possible. 
%For these attacks, the power mode information request, the alive check request, or the \gls*{doip} entity information request messages are used. 

% TLS
A great improvement in the security level, compared to previous versions of the specification and the draft, which were assessed before, poses the addition of encryption and integrity protection using \gls*{tls}. 
%\gls*{tls} provides integrity and confidentiality. 
%Both security properties are therefore fulfilled.
%By using \gls*{tls}, the freshness of messages with the countermeasures is ensured and replay attacks are no longer possible.
%By using \gls*{tls}, consequently non-repudiation is also given, each message is indisputably associated with an entity.
The secure channel provided by \gls*{tls} can mitigate all vulnerabilities present in the diagnostic communication phase, but none except one of the vulnerabilities during vehicle identification and routing activation.
In the standard, the use of \gls*{tls} and authentication is optional for the vehicle manufacturer. 
Consequently, the security of the \gls*{doip} protocol is dependent on the respective vehicle manufacturer and the implementation.
We therefore propose two additional measures to add to the standard:
\gls*{tls} client authentication and a signature over the \gls*{doip} header.

% beyond TLS: authentication/authorization
The ISO\,13400--2~\cite{iso_iso_2019} does not define any authentication procedures.
This is left to the higher layers in the \gls*{osi} model and the vehicle manufacturer. 
%The secrecy and authenticity of the messages can only be ensured when \gls*{tls} is used.
%
To prevent illegitimate access, even when using \gls*{tls}, authentication and authorization is required.
Therefore, \gls*{tls} client authentication should be used to protect the systems from unauthenticated access. 
On application layer, \gls*{uds} could perform an additional, fine-grained authentication mechanism, but a more elegant solution would be to reuse the performed \gls*{tls} authentication for authorization.
The earlier in the protocol phases such an authentication is performed, the more vulnerabilities can be mitigated. 
The earliest vulnerability in the protocol run that \gls*{tls} in \gls*{doip}'s current specification mitigates is the scanning and enumeration of \acrlongpl*{sa} during the routing activation.

% beyond TLS: Signature
The remaining vulnerabilities that are not due to inconsistencies in the standard can be mitigated by an integrity protection during the vehicle identification and routing activation phases by a signature over the header.
The trust in the keys used for the signature can be established from the same \gls*{pki} as the one required for \gls*{tls}, thereby adding no extra effort for key management.
This prevents most non-trivial denial-of-service attacks and the one that exploits the synchronization feature during the vehicle announcement phase.
Confidentiality is not a relevant security property in this phase and since the functionality requires a multicast, \gls*{dtls} for a protection on transport layer is infeasible.
\gls*{ipsec} as proposed by \citet{kleberger_securing_2014} may be an alternative to our proposed signature.
However, we argue that it introduces higher complexity and is redundant in later phases due to the existing \gls*{tls} channel, which was not specified in the standard at the time of \citeauthor{kleberger_securing_2014}'s writing.

% Further recommendations: fingerprinting
In addition to secure design, a protocol's security depends also on the implementation.
Vulnerabilities are inevitably introduced in individual implementations.
Ambiguities must be removed from the standard to prevent that different implementations can easily be discerned and device fingerprinting becomes possible.
When the adversary is able to identify the implementation, they can exploit specific vulnerabilities of this implementation.

%Four components need to be protected: 
%The tester and the vehicle must be protected from intruders and adversaries. 
%\gls*{tls} secures the communication channel between vehicle and tester. 
%\todo{only without tls?!}
%And if the vehicle and the tester are connected via a local area network, such as that of a workshop, then this network must also be secured and protected.

\medskip

We envision that in future work, multiple \gls*{doip} protocol implementations could be verified together with \gls*{tls} in real-world scenarios or in simulations to validate the identified vulnerabilities. 
It should also be investigated whether there are practical limitations of the use of \gls*{tls}, \gls*{tls} client authentication, and signatures that we propose for additional protective measures.
This should account for the limited resources of embedded systems and practical challenges, like key distribution, which potentially restrict the use of \gls*{tls} in \gls*{doip} entities.

\medskip

\gls*{doip} was not designed to be a secure protocol from the start.
\gls*{tls} was added only later and mitigates not all vulnerabilities.
However, \gls*{tls} boosts security to such an extent compared to the bare usage of \gls*{doip} that it should be mandatory together with its client authentication mechanism.
These two measures significantly increase the security of the protocol. 
For additional protection, our proposed measures would need to be adopted in the \gls*{doip} standard first, but they require only small changes to the protocol.
They complement the existing optional \gls*{tls} and authentication measures
and mitigate the remaining vulnerabilities in the phases not covered by the current standard's usage of \gls*{tls}.
%\vfill

\vspace{5ex}
\begin{acks}
The authors thank Gerd Langer, Mercedes-Benz Tech Innovation, for his support for and feedback about our research and Frank Kargl, Ulm University, for his input during the development of our analysis idea.
\end{acks}
\vspace{5ex}
 
%% This section has a special environment:
%% \begin{verbatim}
%%   \begin{acks}
%%   ...
%%   \end{acks}
%% \end{verbatim}
%% so that the information contained therein can be more easily collected
%% during the article metadata extraction phase, and to ensure
%% consistency in the spelling of the section heading.
%% 
%% Authors should not prepare this section as a numbered or unnumbered {\verb|\section|}; please use the ``{\verb|acks|}'' environment.
%% 
%% \section{Appendices}
%% 
%% If your work needs an appendix, add it before the
%% ``\verb|\end{document}|'' command at the conclusion of your source
%% document.
%% 
%% Start the appendix with the ``\verb|appendix|'' command:
%% \begin{verbatim}
%%   \appendix
%% \end{verbatim}
%% and note that in the appendix, sections are lettered, not
%% numbered. This document has two appendices, demonstrating the section
%% and subsection identification method.
%% 
%% 
%% %%
%% %% The acknowledgments section is defined using the "acks" environment
%% %% (and NOT an unnumbered section). This ensures the proper
%% %% identification of the section in the article metadata, and the
%% %% consistent spelling of the heading.
%% \begin{acks}
%% To Robert, for the bagels and explaining CMYK and color spaces.
%% \end{acks}

%%
%% The next two lines define the bibliography style to be used, and
%% the bibliography file.
\bibliographystyle{ACM-Reference-Format}
\bibliography{bibliography}

%%
%% If your work has an appendix, this is the place to put it.
\appendix

%%
%% This section is about the research methods.
%% TODO: Uncomment if we are writing something about the research method.
%
%\section{Research Methods}
%
%\subsection{Part One}
%
%Lorem ipsum dolor sit amet, consectetur adipiscing elit. Morbi
%malesuada, quam in pulvinar varius, metus nunc fermentum urna, id
%sollicitudin purus odio sit amet enim. Aliquam ullamcorper eu ipsum
%vel mollis. Curabitur quis dictum nisl. Phasellus vel semper risus, et
%lacinia dolor. Integer ultricies commodo sem nec semper.
%
%\subsection{Part Two}
%
%Etiam commodo feugiat nisl pulvinar pellentesque. Etiam auctor sodales
%ligula, non varius nibh pulvinar semper. Suspendisse nec lectus non
%ipsum convallis congue hendrerit vitae sapien. Donec at laoreet
%eros. Vivamus non purus placerat, scelerisque diam eu, cursus
%ante. Etiam aliquam tortor auctor efficitur mattis.
%
%\section{Online Resources}
%
%Nam id fermentum dui. Suspendisse sagittis tortor a nulla mollis, in
%pulvinar ex pretium. Sed interdum orci quis metus euismod, et sagittis
%enim maximus. Vestibulum gravida massa ut felis suscipit
%congue. Quisque mattis elit a risus ultrices commodo venenatis eget
%dui. Etiam sagittis eleifend elementum.
%
%Nam interdum magna at lectus dignissim, ac dignissim lorem
%rhoncus. Maecenas eu arcu ac neque placerat aliquam. Nunc pulvinar
%massa et mattis lacinia.

\end{document}